\documentclass{mpe_report}

\usepackage{psfig,graphicx,epsfig}
\usepackage{color}
\usepackage{amsmath,amssymb,epic,eepic,array}

\begin{document}

\pagenumbering{arabic}
\setcounter{page}{128}

\renewcommand{\FirstPageOfPaper }{128}\renewcommand{\LastPageOfPaper }{131}

\title{On the role of the current loss mechanism in radio pulsar evolution}
\author{V.S.Beskin\inst{1} \and E.E.Nokhrina\inst{1,2}}
\institute{P.N.Lebedev Physical Institute, RAS,
             Leninsky prosp., 53, Moscow, 119991, Russia
             \and Moscow Institute of Physics and Technology,
             Institutsky per. 9, Dolgoprudny, 141700, Russia}
\maketitle

\begin{abstract}
The aim of this article is to draw attention to the importance of
the electric current loss in the energy output of radio pulsars.
We remind that even the losses attributed to the magneto-dipole
radiation of a pulsar in vacuum can be written as a result of 
Amp\`ere force action of the electric currents flowing over the
neutron star surface (Michel 1991, Beskin, Gurevich \& Istomin
1993). It is this force that is responsible for the transfer of
angular momentum of a neutron star to an outgoing magneto-dipole
wave. If a pulsar is surrounded by plasma, and there is no
longitudinal current in its magnetosphere, there is no energy
loss (Beskin, Gurevich \& Istomin 1993, Mestel, Panagi \&
Shibata 1999). It is the longitudinal current closing within the
pulsar polar cap that exerts the retardation torque acting on the
neutron star. This torque can be determined if the structure of
longitudinal current is known. Here we remind of the solution by
Beskin, Gurevich \& Istomin (1993) and discuss the validity of
such an assumption. Finally, it is shown that the
behaviour of the recently observed "part-time job" pulsar B1931+24
can be naturally explained within the model of current loss while
the magneto-dipole model faces difficulties.
\end{abstract}

\section{Introduction}

The recent observations of "part-time job" pulsars (Kramer et al 2006)
such as pulsar B1931+24 has drawn attention to the particular
mechanism of the energy losses of pulsars. In this article we 
summarize some results obtained for the model of current
losses (Beskin, Gurevich \& Istomin 1993) and the consequences of it.

Radio pulsars are definitely nonaxisymmetric objects. However,
the most results both in electrodynamics of the pulsar
magnetosphere and in neutron star statistics were obtained under
the assumption that the magnetic axis is parallel to the
rotational one. Taking into account the inclination angle one can
change qualitatively the consequences of the standard model.
E.g., one can show that for the orthogonal rotator with the local
GJ longitudinal electric current the light surface (where
$|{\textbf E}|^2 = |{\textbf B}|^2$) must be located in the very
vicinity of the light cylinder. In this case it is impossible to
prolong the MHD flow up to infinity, so the effective energy
conversion and the current closure is to take place near the
light surface.

\section{Magneto-dipole loss}
At first, the energy loss mechanism of radio pulsars has been 
connected with the magneto-dipole radiation~(Pacini 1967). Indeed,
the magneto-dipole radiation power
\begin{equation}
W_{{\mathrm {md}}}
=\frac{1}{6}\frac{B_{0}^{2}\Omega^{4}R^{6}}{c^{3}}\sin^{2}{\chi},
\label{MD-0}
\end{equation}
where $\chi$ is the angle between rotational and magnetic axis,
$R \sim 10$ km is a neutron star radius, and $\Omega$ is a
pulsar angular velocity explains pulsar activity
and observed energy loss for expected large magnetic field near
the surface $B_{0}\sim 10^{12}$ G.

Let us recall that the physical reason of such energy loss is the
action of the torque exerted on the pulsar by the Amp\`ere force of
the electric currents flowing over the neutron star surface
(Istomin 2005). The electric and magnetic fields in the outgoing
magneto-dipole wave in vacuum can be found by solving the wave
equations $\nabla^2\mathbf{B}+{\Omega^{2}}/{c^{2}}\mathbf{B}=0$
and $\nabla^2\mathbf{E}+{\Omega^{2}}/{c^{2}}\mathbf{E}=0$ with
the boundary conditions stated as the fields corresponding
components $\mathbf{E}_{\mathrm t}$ and $\mathbf{B}_{\mathrm n}$
being continuous through the neutron star surface. Inside the
star one can consider magnetic field as homogeneous, and find the
corresponding electric field using the frozen-in condition. As a
result, the discontinuity of $\{\mathbf{B}_{\mathrm t}\}$ and 
$\{\mathbf{E}_{\mathrm n}\}$ give us the surface charge 
$\sigma_{\mathrm s}$ and the surface current
\begin{equation}
\mathbf{J}_{\mathrm s}=\frac{c}{4\pi}[\mathbf{n},\{\mathbf{B}\}].
\end{equation}
The Amp\`ere force exerts the torque
\begin{equation}
\mathbf{K}=\frac{1}{c}\int{[\mathbf{r},[\mathbf{J}_{\mathrm
s},\mathbf{B}]]{\rm d}S}
\end{equation}
on the neutron star. The energy loss of a pulsar due to this
torque is equal to (\ref{MD-0}). Thus, it is the surface current
that is responsible for the angular momentum transform from a
neutron star to an outgoing magneto-dipole wave (Michel 1991,
Beskin, Gurevich \& Istomin 1993).

Thus, a pulsar in vacuum loses its rotational energy due to angular
momentum transform to the electromagnetic wave at the rate given
by (\ref{MD-0}). However, this is not so if the pulsar magnetosphere 
is filled with plasma and there is no longitudinal current in the 
magnetosphere. As was shown (Beskin, Gurevich \& Istomin 1993, Mestel, 
Panagi \& Shibata 1999), in this case the Poynting
flux through the light cylinder is equal to zero. Indeed, as the
ideal conductivity condition is applicable not only inside the
neutron star but outside as well there is no magnetic field
discontinuity at the star surface. Consequently, there is no
Amp\`ere force acting on a pulsar and, hence, there is no energy
loss. For zero longitudinal current the light cylinder is a
natural boundary of the pulsar magnetosphere.

\section{Current loss}
In this section we remind the exact solution for the surface
current within the polar cap presented in the monograph by
Beskin, Gurevich \& Istomin (1993). 
As it was shown in the previous section, the neutron star retardation 
is due to Amp\`ere force ${\bf F}_{\rm A} = {\bf J}_{\rm s} \times
{\bf B}/c$. If the magnetosphere is filled with plasma, the
surface current ${\bf J}_{\rm s}$ is flowing within magnetic polar
cap only. This surface current closes the volume longitudinal
current in the magnetosphere and the return current flowing along
the separatrix between open and closed field lines region.

In order to write the equation for the surface current, the
several assumptions must be made. We assume that the conductivity
of the pulsar surface is uniform, and the electric field ${\bf
E}_{\rm s}$ has a potential, so that the surface current can be
written as ${\bf J}_{\rm s} = {\bf {\nabla}}\xi'$. Using the
stationary continuity equation ${\rm div}{\bf J}=0$ where
$\partial J_{z}/\partial z$ is equal to the volume current
$i_{||}B_{0}$ flowing along the open field lines one can obtain
\begin{equation}
\nabla_{(2)}^2 \xi' = -i_{||}B_{0}. \label{6}
\end{equation}

Making the substitution $x=\sin{\theta_m}$ and introducing the
non-dimensional potential $\xi=4\pi\xi'/B_{0}R^{2}\Omega$ and
current $i_{0}=-4\pi i_{||}/\Omega R^{2}$ we get
\begin{equation}
\left(1-x^{2}\right)\frac{\partial^{2}\xi}{\partial
x^{2}}+\frac{1-2x^{2}}{x}\frac{\partial\xi}{\partial
x}+\frac{1}{x^{2}}\frac{\partial^{2}\xi}{\partial\varphi_{m}^{2}}
= i_{0}(x,\varphi_{m}). \label{7}
\end{equation}
Here $\theta_m$ and $\varphi_{m}$ are polar and azimuth angles
with respect to magnetic axis.

Equation (\ref{7}) needs a boundary condition. This boundary
condition results from the proposition that there is no surface
current outside the magnetic polar cap. This means that
\begin{equation}
\xi\left[x_{0}(\varphi_{m}),\varphi_{m}\right] = {\rm const},
\label{a1}
\end{equation}
where $x_{0}(\varphi_{m})$ is the polar cap boundary. Indeed, let
us suppose that there is no such boundary condition for the
potential $\xi$. In this case we get $\nabla_{(2)}^2\xi=r.h.s$ with
$r.h.s.=i_{0}$ inside the polar cap and $r.h.s=0$ outside it. The
solution of homogeneous equation is
\begin{equation}
\left.\xi(x,\varphi_{m})\vphantom{\tilde|}\right|_{x\geq
x_0}=\sum_{n=0}^{\infty}\left(\frac{1-x}{1+x}\right)^{n/2}
f_{n}(\varphi_{m}), \
\end{equation}
\begin{equation}
\left.\xi(x,\varphi_{m})\vphantom{\tilde|}\right|_{x < x_0}=
\sum_{n=0}^{\infty}\left(\frac{1+x}{1-x}\right)^{n/2}
f_{n}(\varphi_{m}),
\end{equation}
where $f_{n}(\varphi_{m})=
(a_{n}\cos{n\varphi_{m}}+b_{n}\sin{n\varphi_{m}})$ with $a_{n}$
and $b_{n}$ being arbitrary constants. In this case the surface
current ${\bf J}_{\rm s} \propto \nabla\xi$ is circulating over
the whole neutron star surface resulting in arbitrary energy
loss. However, this means that there is a potential drop between
different points of a neutron star surface which inevitably leads
to the volume current in the region of closed field lines in the
magnetosphere. This contradicts to the assumption that there are
no longitudinal currents flowing in the region of closed field lines. 
Thus, the boundary condition (\ref{a1}) is to be postulated. In this 
case the jump in the potential derivative \{$\nabla \xi$\} at
$x=x_{0}(\varphi_{m})$ gives us the current flowing along the
separatrix. As we see, it is defined uniquely by the longitudinal
current in the region of open field lines and by condition that no
longitudinal current can flow in the region of the closed field
lines.

For arbitrary inclination angle $\chi$ the electric current $i_{0}$ 
can be written as a sum of its symmetric $i_{\rm S}$ and anti-symmetric
$i_{\rm A}$ components. The anti-symmetric current begins playing
the main role when the pulsar polar cap crosses the surface where
the Goldreich-Julian charge density $\rho_{\rm GJ}=-{\bf
\Omega}\cdot{\bf B}/2\pi c$ changes sign. This condition can be
written as
\begin{equation}
\chi=\frac{\pi}{2}-\sqrt{\frac{\Omega R}{c}}.
\end{equation}
For example, taking the Goldreich-Julian current density
\begin{equation}
i_{\rm GJ}(x,\varphi_{m}) \approx \cos{\chi} + \frac{3}{2} x
\cos\varphi_{m} \sin{\chi}=i_{\rm S}+i_{\rm A}x\cos\varphi_{m},
\label{8}
\end{equation}
(and assuming $x_0 = $ const)
we obtain the following solutions of the Dirichlet problem
(\ref{7})--(\ref{a1}) for the symmetric and anti-symmetric volume
currents respectively:
\begin{equation}
\xi_{\rm S}=\frac{i_{\rm S}}{4}x^{2},
\end{equation}
\begin{equation}
\xi_{\rm A}=\frac{i_{\rm
A}}{8}x(x^{2}-x_{0}^{2})\cos{\varphi_{m}}.
\end{equation}

The torque exerted by the surface current over the neutron star
can be written as
\begin{equation}
{\bf K} = \frac{1}{c}\int \left[{\bf r}, \left[{ \bf J}_{s},
\left({\bf B}_{0}\right)\right]\right]{\rm d}S, \label{9}
\end{equation}
where $\bf B_{0}$ is the dipole field near the neutron star
surface. Let us expand the braking torque ${\bf K}$ over the
orthogonal system of unit vectors ${\bf e}_{m}$, ${\bf n}_{1}$,
and ${\bf n}_{2}$. Here ${\bf e}_{m}$ is a unit vector along the
magnetic moment ${\bf m}$; vector ${\bf n}_{1}$ is perpendicular 
to the magnetic moment and lies in the plane of the magnetic moment 
and the rotational axis; vector ${\bf n}_{2}$ complements these to
the right-hand triple:
\begin{equation}
{\bf K}=K_{||}{\bf e}_{m}+K_{\bot}{\bf n}_{1} +K_{\dag}{\bf
n}_{2}. \label{10}
\end{equation}
$K_{\dag}$ plays no role in Euler equations that describe the
rotational dynamics of the decelerating neutron star. As a result
we have (Beskin, Gurevich \& Istomin 1993)
\begin{equation}
K_{||}= -\frac{B_{0}R^{2}}{c} \int_{0}^{2\pi}{\rm
d}\varphi_{m}\int_{0}^{x_{0}(\varphi_{m})}{\rm d}x\,
x^{2}\sqrt{1-x^{2}}\frac{\partial\xi}{\partial x},
\label{11}
\end{equation}
\begin{equation}
K_{\perp}=K_{1}+K_{2}, \label{12}
\end{equation}
where
\begin{eqnarray}
K_{1} & = & \frac{B_{0}R^{2}}{c}\int_{0}^{2\pi}{\rm d}\varphi_{m}
\int_{0}^{x_{0}(\varphi_{m})}{\rm d}x A,
\label{13} \\
K_{2} & = & \frac{B_{0}R^{2}}{c}\int_{0}^{2\pi}{\rm d}\varphi_{m}
\int_{0}^{x_{0}(\varphi_{m})}{\rm d}x\,
x^{3}\cos{\varphi_{m}}\frac{\partial\xi}{\partial x}, \label{14}
\end{eqnarray}
and $\displaystyle A=x\cos{\varphi_{m}}{\partial\xi}/{\partial
x} -\sin{\varphi_{m}}{\partial\xi}/{\partial\varphi_{m}}$.

As one can find (Beskin, Gurevich \& Istomin 1993), the torque component 
$K_{1}$ is equal to zero equivalently for arbitrary shape of the polar 
cap due to the boundary condition (\ref{a1}). Thus, the values $K_{||}$ 
and $K_{\perp}$ can be written as
\begin{eqnarray}
K_{||} & = &
\frac{B_{0}^{2}\Omega^{3}R^{6}}{c^{3}}\left[-c_{||}i_{\rm S}
-\mu_{||}\left(\frac{\Omega R}{c}\right)^{1/2}i_{\rm A}\right],
\label{16} \\
K_{\perp} & = & \frac{B_{0}^{2}\Omega^{3}R^{6}}{c^{3}}
\left[\mu_{\perp}\left(\frac{\Omega R}{c}\right)^{1/2}i_{\rm S}
+c_{\perp}\left(\frac{\Omega R}{c}\right)i_{\rm A}\right].
\label{17}
\end{eqnarray}
Here the coefficients $\mu_{||}$ and $\mu_{\perp}$ depending on
the shape of the polar cap are much less than unity, and the
coefficients $c_{||}$ and $c_{\perp}\sim 1$.

We can now find the derivatives of the angular velocity
$\dot{\Omega}$ and of the inclination angle $\dot{\chi}$ of a
neutron star through the Euler dynamics equations:
\begin{eqnarray}
J_r\frac{{\rm d}\Omega}{{\rm d}t} & = &
K_{||}\cos{\chi}+K_{\perp}\sin{\chi},
\label{18} \\
J_r\Omega\frac{{\rm d}\chi}{{\rm d}t} & = &
K_{\perp}\cos{\chi}-K_{||}\sin{\chi}. \label{19}
\end{eqnarray}
For the inclination angles $\chi$ not too close to $90^{\circ}$
(i.e., for $\cos\chi> (\Omega R/c)^{1/2}$) when the
anti-symmetric current plays no role in the neutron star dynamics
we obtain
\begin{eqnarray}
\label{Losses1} \frac{{\rm d}\Omega}{{\rm d}t} & = &
-c_{||}\frac{B_{0}^{2}\Omega^{3}R^{6}}{J_r c^{3}}i_{\rm
S}\cos{\chi},
\label{20} \\
\frac{{\rm d}\chi}{{\rm d}t} & = &
\phantom{-}c_{||}\frac{B_{0}^{2}\Omega^{2}R^{6}}{J_r c^{3}}i_{\rm
S}\sin{\chi}. \label{21}
\end{eqnarray}

As a result, for homogeneous current density within open magnetic
field line region $i_{\mathrm S} = j_{||}/j_{\mathrm GJ} =$ const
where $j_{\mathrm GJ}= c \rho_{\mathrm GJ}$ we have
\begin{equation}
W_{\mathrm c} = \frac{f_*^2}{4} \,
\frac{B_{0}^{2}\Omega^{4}R^{6}}{c^{3}} i_{\mathrm S}\cos{\chi}.
\label{c-0}
\end{equation}
Here $f_{*}$ is the non-dimensional area of a pulsar polar cap:
$S_{\rm cap}=f_{*}\pi(\Omega R/c)R^2$. It depends on the structure
of the magnetic field near the light cylinder. For a pure dipole
magnetic field (and aligned rotator) $f_* = 1$, and for a
magnetosphere containing no longitudinal currents $f_*$ changes
from $1.592$ for the aligned rotator, $\chi=0^\circ$ (Michel 1991),
to $1.96$ for an orthogonal rotator, $\chi = 90^{\circ}$
(Beskin, Gurevich \& Istomin 1993). Recent numerical calculations for an
axisymmetric magnetosphere with non-zero longitudinal electric
current give $f_* \approx 1.23$ -- $1.27$~(Gruzinov 2005,
Komissarov 2006, Timokhin 2006). If the singular
point separating open and close field lines can be located inside
the light cylinder, the value $f_*$ can be even $\gg 1$. As the
Goldreich-Julian charge density near the polar cap is proportional
to $\cos\chi$, one can write
\begin{equation}
W_{\mathrm c} \approx \frac{f_*^2}{4} \,
\frac{B_{0}^{2}\Omega^{4}R^{6}}{c^{3}} \cos^2{\chi}. 
\label{c-1}
\end{equation}

On the other hand, for $\chi \approx 90^{\circ}$ when the
anti-symmetric current plays the leading role we obtain
\begin{equation}
W_{\mathrm c} \approx \frac{B_{0}^{2}\Omega^{5}R^{7}}{c^{4}} \, i_{\rm A}.
\label{c-2}
\end{equation}
As we see, the energy loss of the orthogonal rotator is 
$\Omega R/c$ times smaller than that of the aligned rotator.

\section{PSR B1931+24}
The recent discovery of the "part-time job" pulsar PSR B1931+24
(Kramer et al 2006) with $\dot\Omega_{\rm on}/\dot\Omega_{\rm off}
\approx 1.5$ shows that the current loss is indeed playing an
important role in the pulsar energy loss. If we assume that in
the on-state the energy loss is connected with the current loss
only and in the off-state with the magneto-dipole radiation (in
which case the magnetosphere must be not filled with plasma) we
get
\begin{equation}
\frac{\dot\Omega_{\rm on}}{\dot\Omega_{\rm off}} = \frac{3
f_*^2}{2} \, {\rm cot}^2\chi.
\end{equation}
It gives $\chi \approx 60^{\circ}$. On the other hand,
%
if we assume the Spitkovsky's relation for the on-state energy
loss (Spitkovsky 2006)
\begin{equation}
W_{\rm tot} = \frac{1}{4} \, \frac{B_0^2 \Omega^4 R^6}{c^3} (1 +
\sin^2\chi),
\end{equation}
we obtain
\begin{equation}
\frac{\dot\Omega_{\rm on}}{\dot\Omega_{\rm off}} = \frac{3}{2} \,
\frac{(1+ \sin^2\chi)}{\sin^2\chi}.
\end{equation}
Clear, this ratio cannot be equal to 1.5 for any inclination
angle. This discrepancy can be connected with that fact that all
the numerical calculations produced recently contain no
restriction on the longitudinal electric current. As a
result, current density can be much larger than Goldreich-Julian
one.

\section{On the magnitude of a surface current}
As we have shown, the current loss plays the major role in the
pulsar dynamics. In particular, the behaviour of the pulsar
B1931+24 can be naturally explained within this model. The
current loss model have some important consequences:
\begin{enumerate}
\item the energy loss of an orthogonal rotator is $\Omega R/c$
times smaller than for an aligned rotator. This is connected with
the boundary condition (\ref{a1}) that leads to almost full
screening of the toroidal magnetic field in the open field lines
region (see Beskin \& Nokhrina 2004);
\item consequently, during its evolution a pulsar inclination
angle tends to $\pi/2$ where energy loss is minimal.
\end{enumerate}
On the other hand, it is known for the Michel's monopole solution
that in order to have the MHD flow up to infinity the toroidal
magnetic field must be of the same order as the poloidal electric
field on the light cylinder. If the longitudinal current $j_{||}$
does not exceed by $(\Omega R/c)^{-1/2}$ times the respective
Goldreich-Julian current density (for the typical pulsars this
factor approach the value of $10^2$), the light surface
$|{\mathbf{E}}| = |\mathbf{B}|$ for the orthogonal rotator must
be located in the vicinity of the light cylinder. In this case
the effective energy conversion and the current closure are to
take place in the boundary layer near the light surface
(Beskin, Gurevich \& Istomin 1993, Chiueh, Li \& Begelman 1998, 
Beskin \& Rafikov 2000).

In order for these results being not true (for example, in order
for the light surface being removed to infinity)
there must be a sufficient change in the current density value in
the inner gap. We should emphasize that for the model with free
particle escape it is hard to support the current different than
the Goldriech-Julian one. Indeed, since $\rho_{\rm GJ}$ is
the particle density needed to screen the longitudinal electric
field, the current  $j_{||}$ must be close $j_{\rm
GJ}=c\rho_{\rm GJ}$. In order to change this value significantly
we must support the plasma inflow in the inner gap region
(Lyubarskii 1992). For example, these particles can be produced in
the outer gap. But it is obvious that for any poloidal field 
configuration the major number of field lines intersect the
outer gap region inside the Alfv\'enic surface. 
Inside the Alfv\'enic surface the flow remains still highly 
magnetized. Thus, the deviation of current lines from the field 
lines is negligible. On the other hand, magnetized plasma can 
intersect the Alfv\'enic surface outwards only (see Beskin 2006 
for more detail). Thus, the outer gap can not significantly affect 
the current in the vicinity of the polar cap.

Finally, it is necessary to stress that the recent numerical
calculations (Gruzinov 2005, Komissarov 2006, Timokhin 2006, 
Spitkovsky 2006) do not include into consideration that the 
longitudinal current density must be close to $j_{\rm GJ}$. 
In all these works the authors assume that the current flowing 
through the cascade region can be arbitrary. However, if this 
is not so, and the current is indeed close to the Goldreich-Julian 
current, the structure of a magnetosphere may be different from 
the ones obtained in the numerical simulations.

\section{Conclusion}
As we have seen, the current loss connected with the longitudinal
current flowing in the magnetosphere plays the main role in pulsar
dynamics, and recent observations of "part-time job" pulsar
supports this point. This evolution includes not only a neutron
star retardation but also the sufficient change in the angle
between magnetic and spin axis. We have seen as well that the
model of current loss depends crucially on the distribution of the
electric current and its value in the inner gap. For current loss
model  the inclination angle grows with time so a
pulsar tends to be an orthogonal rotator. In this case the energy
loss is to be $\Omega R/c$ times smaller than for the aligned
rotator. As a consequence, the light surface must be located in
the very vicinity of the light cylinder. On the other hand, to
realize the homogeneous MHD outflow up to infinity for the
orthogonal rotator the current density in the open field line
region is to be much larger than $j_{\rm GJ}$.

\begin{acknowledgements}
The authors thank A.V.Gurevich and Ya.N.Istomin for the useful 
discussions. This work was partially supported by the Russian 
Foundation for Basic Research (Grant no.~05-02-17700) and Dynasty 
fund.
\end{acknowledgements}

            \clearpage

\end{document}